# Deep-Learning Assisted Compact Modeling of Nanoscale Transistor

**Hei Kam**
Stanford University
heikam2112@gmail.com

**Abstract**

Transistors are the basic building blocks for all electronics. Accurate prediction of their current-voltage (*IV*) characteristics enables circuit simulations before the expensive silicon tape-out. In this work, we propose using deep neural network to improve the accuracy for the conventional, physics-based compact model for nanoscale transistors. Physics-driven requirements on the neural network are discussed. Using finite element simulation as the input dataset, together with a neural network with roughly 30 neurons, the final IV model can well-predict the IV to within 1%. This modelling methodologies can be extended for other transistor properties such as capacitance-voltage (*CV*) characteristics, and the trained model can readily be implemented by the hardware description language (HDL) such as VerilogA for circuit simulation.

## 1. Introduction

Transistors are four-terminal switches that form the basic building blocks for all electronics. Over the past 50 years, transistor scaling driven by the Moore's Law has led to tremendous improvement in integrated circuit performance. A significant part of this was enabled by physics-based transistor model which allows for efficient circuit design and simulation (**Fig.1**).

Transistor current-voltage (*IV*) models are a set of equations that predicts the drain-to-source current $I_{DS}$ based upon the input gate/source/drain voltages ($V_G$, $V_S$, $V_D$) (**Fig. 2**). These models were traditionally derived based upon physics, but close-form solutions that accounts for nanoscale effects are either mathematically complicated or do not exist. In this work, we propose using neural network for accuracy improvement. Such neural network can easily be extended for other device parameters and implemented in hardware description language such as *VerilogA* for very large-scale integration (VLSI) circuit design and simulation.

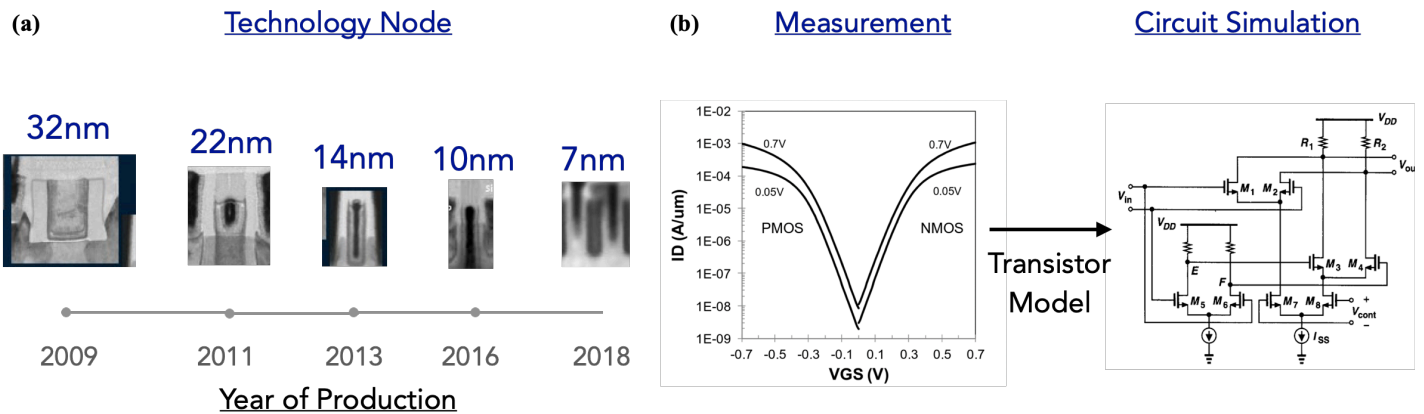

**Fig. 1 (a)** TEM images showing transistor technology since 2009. Following Moore's Law, transistor feature size continued to scaled. **(b)** Measured IV for Intel 14nm transistor technology. Utilizing transistor models, circuit designers simulate and predict circuit performance with fast turnaround.

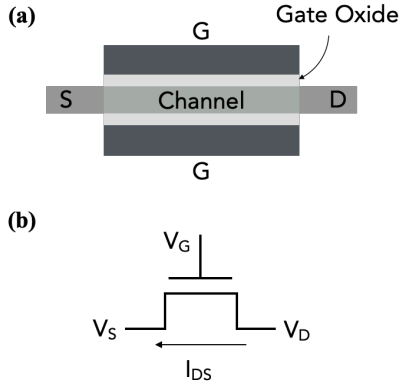
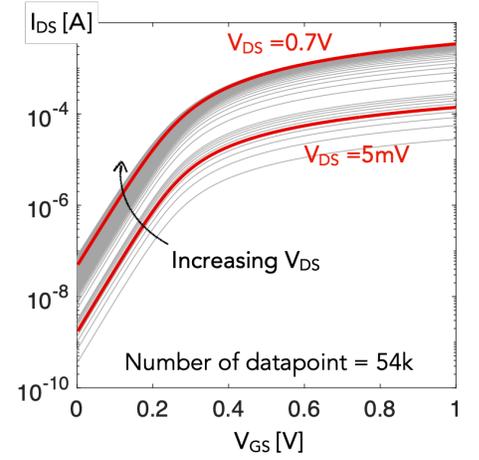

**Fig. 2.** (a) Cross-section schematic and (b) circuit element for a FinFET. $I_{DS}$ can be expressed either as a function of ($V_{GS}$,$V_{DS}$) or ($V_{GS}$,$V_{GD}$).

**Table I** Device parameter for FEM simulation. $V_{GS}$ and $V_{DS}$ range from -0.5V to 1V and 0V to 1V, respectively.

**Fig. 3.** Simulated $I_{DS}$-$V_{GS}$ characteristics.

## 2. Related Work

We begin our discussion with the Enz-Krummenacger-Vittoz (EKV) Core IV Model[1] [1], in which $I_{DS}$ for an ideal long-channel transistor can be modeled by 4 device parameters ($P$, $V_{SS}$, $V_T$, $\beta$)[2]:

$$I_{DS,m}(V_{GS}, V_{GD}) \approx P[\phi(V_{GS})^\beta - \phi(V_{GD})^\beta] \;,\; \phi(V_{GX}) = V_{ss} \log\left(1 + \exp\left(\frac{V_{GX}-V_T}{V_{ss}}\right)\right) \tag{1}{}^{3}$$

where $P$ is the prefactor that is set by the device dimension and mobility, subscript X denotes source (S) or drain (D). $V_{SS}$ is related to the thermal voltage, $V_T$ is the threshold voltage, and $\beta = 2$.

We first compare the core IV model against the IV characteristics simulated by finite element method **(Fig.3-4)** using commercially available Poisson-drift-diffusion solver. Device parameters are tabulated in **Table I**. As shown in **Fig. 4**, this simple model well-matches to device at low $V_{DS}$; but its accuracy decreases at high voltage biases. In conventional approach, analytical models that account for non-idealities such as series resistance, drain-induced barrier lowering (DIBL), channel length modulation [1] are added to improve the accuracy. However, many of these models involve solving nonlinear differential equation, and close-form solutions do not exist. By sacrificing some degree of accuracy, empirical models with fitting parameters are commonly used. Recently, pure look-up-table [4] or deep-learning based transistor models [5-10] have been proposed, in which the measured *IV* data were tabulated or used to train the model. As will be discussed in the next section, this approach can potentially result in non-physical behavior and violate fundamental law of physics.

## 3. Physics-driven Requirements for the Transistor Model

The approach we propose herein combines the advantages of both physics-based and deep-learning-based modeling strategies. To this end, we introduce a bias-dependent correction function $\varepsilon(V_{GS},V_{GD})$ to account for the non-idealities:

$$I_{DS}(V_{GD}, V_{GS}) = I_{DS,m}(V_{GS}, V_{GD}) \times \varepsilon(V_{GS}, V_{GD}) = P[\phi(V_{GS})^\beta - \phi(V_{GD})^\beta] \times \varepsilon(V_{GS}, V_{GD}) \tag{2}$$

in which $\varepsilon$ is to be trained by the neural network. Before doing so, we first discuss the physics-driven requirements for $I_{DS}$ and $\varepsilon$.

---

[1] Note that the EKV model is used as an example. Other transistor models such as BSIM-MG [2] or PSP [3] model can also be used.
[2] Source-drain current $I_{DS}$ is set by gate/drain/source voltages and can be expressed as a function of ($V_{GS}$,$V_{DS}$) or ($V_{GS}$,$V_{GD}$). We will use them interchangeably in this work. $V_{XY}=V_X-V_Y$ is the voltage drop across node X and Y, which can be gate G, drain D or source S.
[3] Fun fact: First derivative of function f(x)=log(1+exp(x)) is the sigmoid function, and it resembles the *ReLU* function without the gradient discontinuity at *x*=0.

According to the Ohm's Law, $I_{DS}$ must be zero when $V_{DS}=0$. Many deep-learning-based transistor models fail to predict so and violate the Ohm's Law [8-10]. The advantage of using the Eq. (2) is that $I_{DS,m}$ is always zero at $V_{DS}=0$ and it therefore satisfies the requirement.

Furthermore, since transistors are symmetric devices, the direction of the current flow change if we swap the source and drain voltage, i.e. $I_{DS}(V_{GS}, V_{GD}) = -I_{DS}(V_{GD}, V_{GS})$. Substituting this condition into (2), we get:

$$P[\phi(V_S)^\beta - \phi(V_D)^\beta] \times \varepsilon(V_{GD}, V_{GS}) = -P[\phi(V_D)^\beta - \phi(V_S)^\beta] \times \varepsilon(V_{GS}, V_{GD})$$

i.e. the correction function must be symmetric:

$$\varepsilon(V_{GD}, V_{GS}) = \varepsilon(V_{GS}, V_{GD}) \qquad (3)$$

This condition can be satisfied if we preprocess input ($V_{GS}, V_{GD}$) using the following transformation $T$ (**Fig. 5**):

$$T(V_{GS}, V_{GD}) = (V_{GS} + V_{GD}, (V_{GS} - V_{GD})^2) = (V_{GS} + V_{GD}, V_{DS}^2) \qquad (4)$$

To prove that $T$ a is symmetric, we swap the source and drain voltages in Eq. (4), i.e. $T(V_{GD}, V_{GS}) = (V_{GD} + V_{GS}, V_{SD}^2) = (V_{GS} + V_{GD}, V_{DS}^2) = T(V_{GS}, V_{GD})$. Therefore condition (3) is satisfied.

Finally, $I_{DS}$ must be infinitely differentiable with respect to $V_{GS}$ and $V_{DS}$. As a result, the correction function and therefore, the activation function must be infinitely differentiable. In addition, analog circuit applications also require accurate prediction for both the transconductance ($dI_{DS}/dV_{GS}$) and the output conductance ($dI_{DS}/dV_{DS}$). With these requirements, we will discuss the input dataset and choice of network architecture in the next section.

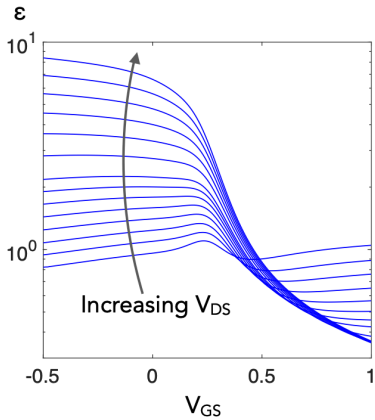

**Fig. 4.** Correction function $\varepsilon \equiv I_{DS}/I_{DS,m}$ Accuracy of the core model decreases at high $V_{DS}$, as indicated by $\varepsilon \gg 1$.

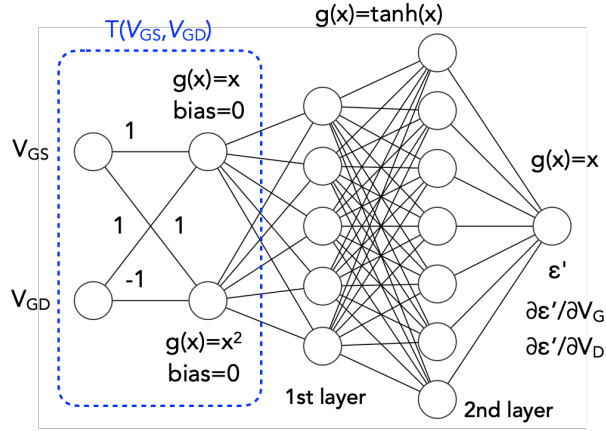

**Fig. 5.** Neural network used in this work. Input is first transformed by $T$ before feeding into the 3-layer network. Transformation $T(V_{GS}, V_{GD}) = (V_{GS}+V_{GD}, V_{DS}^2)$

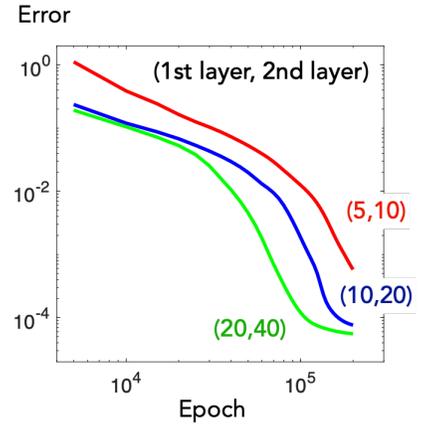

**Fig. 6.** Error vs epoch for various neural network design. Error decreases with increasing number of neurons, as expected.

## 4. Dataset and Feature

Ideally, we prefer to use the measured data from the silicon foundry as the dataset. For the purpose of this project, it suffices to use the finite element method to generate the *IV* data. **Fig 2** shows the simulated transistor structure based upon Intel 14nm FinFET [4] transistor technology [5] [11]. Semiconductor models such as concentration dependent mobility, parallel field mobility and Shockley-Read- Hall (SRH) recombination are included. Quantum tunneling model for gate leakage current is ignored. As shown in **Fig.3**, A dataset of ($V_{GS}$, $V_{DS}$, $I_{DS}$) point with $V_{GS}$ range from -0.5 to 1V and $V_{DS}$ ranges from 1mV to 1V is simulated. Since the $V_{GS}$ and $V_{DS}$ ranges are similar, they are not normalized. By definition, $\varepsilon$ is the normalized $I_{DS}$:

---

[4] FinFET stands for Fin-type field-effect transistor, a self-aligned double-gate transistor that has been in mass production since 2011.
[5] Note that technology node name (e.g. "14nm") used to represent the critical feature size of a transistor. In recent years, however, it has become a marketing term and no longer has physical meaning. Physical gate length $L$ for recent transistor is roughly 20-25nm.

$$\varepsilon(V_{GS}, V_{GD}) \equiv \frac{I_{DS}(V_{GD}, V_{GS})}{I_{DS,m}(V_{GD}, V_{GS})} = \frac{I_{DS}(V_{GD}, V_{GS})}{P[\phi(V_S)^\beta - \phi(V_D)^\beta]}$$

$\partial\varepsilon/\partial V_G$ and $\partial\varepsilon/\partial V_D$ are also computed from the data using the finite difference method. The dataset ($V_{GS}$, $V_{GD}$, $I_{DS}$, $\varepsilon$, $\partial\varepsilon/\partial V_G$, $\partial\varepsilon/\partial V_D$) will be used as the input data to train the neural network.

## 5. Methods

**Fig. 5** shows the architecture for the 3-layer neural network together with the aforementioned transformation *T*. Hyperbolic tangent function *tanh* is used as the activation function for the input and hidden layers due to its infinite differentiability. The cost function J is the mean square error:

$$J = \frac{1}{m}\sum_{i=1}^{m}\left[\left(\varepsilon^{(i)} - \varepsilon'^{(i)}\right)^2 + \eta_G\left(\frac{\partial\varepsilon^{(i)}}{\partial V_G} - \frac{\partial\varepsilon'^{(i)}}{\partial V_G}\right)^2 + \eta_D\left(\frac{\partial\varepsilon^{(i)}}{\partial V_D} - \frac{\partial\varepsilon'^{(i)}}{\partial V_D}\right)^2\right] \quad (5)$$

where *m* is the sample size, first to third terms correspond to the cost in $I_{DS}$, transconductance ($\sim\partial\varepsilon'/\partial V_G$) and output conductance ($\sim\partial\varepsilon'/\partial V_D$), respectively. $\eta_G$ and $\eta_D$ are scaling factors for the gradient costs and ($\eta_G$, $\eta_D$) = (0.5, 1E-3) is used to strike the right balance between various costs. Detailed derivation for the predicted $\varepsilon$ (denoted as $\varepsilon'$), its derivatives $\partial\varepsilon'/\partial V_G$, $\partial\varepsilon'/\partial V_D$ and Eq. (5) is included in the Appendix. The 12k training dataset is purposely selected to cover the full ($V_{GS}$, $V_{GD}$) range of 0 to 0.7V[6], with particular focus on low ($V_{DS}$, $V_{GS}$) for denser sampling at low $I_{DS}$ regime. A larger test set of 56k datapoint with very fine increment of $V_{GS}$, $V_{GD}$ is used, leading to a training-to-test-set ratio of roughly 20/80. TensorFlow with adaptive moment estimation ("Adam") optimization is used for training. Adam optimization is a gradient-descent-based optimization algorithm in which the per parameter learning rate is estimated based upon the first and second moments of the gradients [12]. Given the small training size of < 20000, a single batch is feed into the neural network.

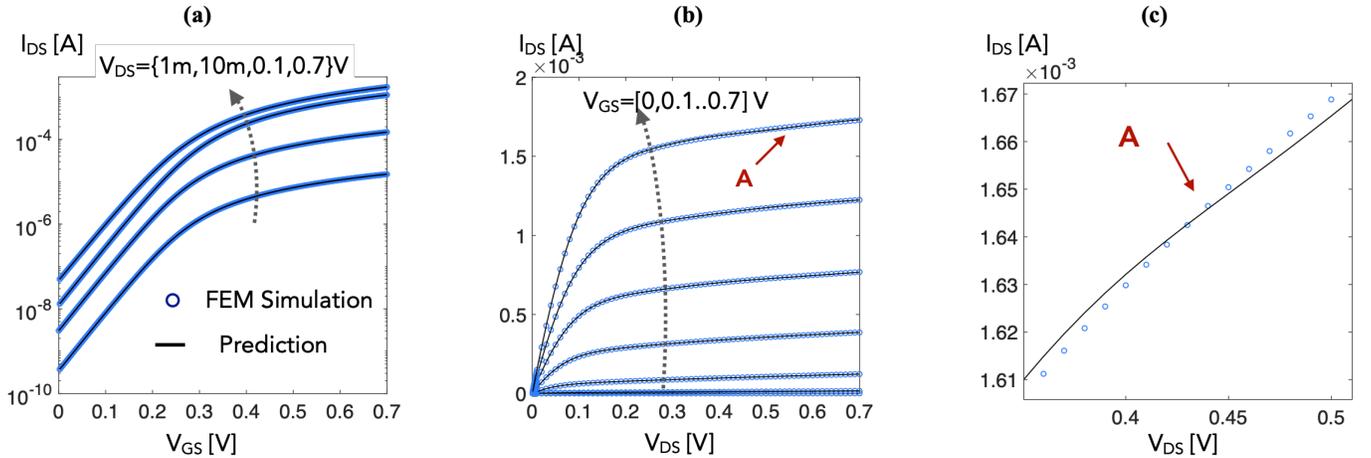

**Fig. 7.** FEM simulated vs neural-network predicted (a) $I_{DS}$-$V_{GS}$ and (b) $I_{DS}$-$V_{DS}$ characteristics. (c) Even though the prediction well-matches to the simulation to within 1%, the output conductance ($dI_{DS}/dV_{DS}$) can deviate up to 20% from simulation at high $V_{GS}$ and $V_{DS}$ (e.g. point A).

## 6. Results and Discussion

**Fig 6.** shows cost vs epoch for various network design. As an example, a trained neural network with 2 hidden layers of 10 and 20 neurons well predicts the correction function to within 1%, and the deep learning assisted model accurately predicted the simulated *IV* and its derivatives (**Fig. 7 and 8**). In general, accuracy improves with increasing number of neurons, as expected. This tunability allows the circuit designers to make direct tradeoffs between model complexity and circuit simulation speed. For example, models with different accuracies can be trained, in which the more complex model is selectively used for the few critical transistors in the circuit.

Also, our discussion thus far focuses only on the EKV model in its simplest form. As technology progresses and better transistor models become available, device engineers can update and retrain the correction function (using Eq (2)). This can potentially simplify the neural network and reduce the number of model parameters.

---

[6] Supply voltage ($V_{DD}$) for Intel 14nm technology is 0.7V, which set the voltage range limit.

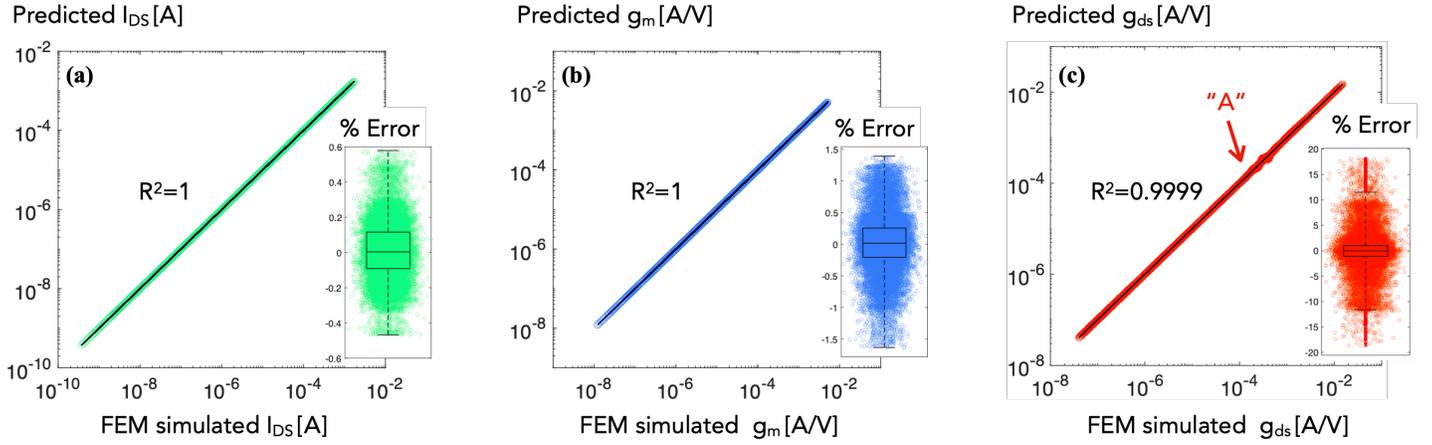

**Fig. 8.** Predicted vs FEM simulated values and the % error distribution for **(a)** current $I_{DS}$ **(b)** transconductance $g_m$ and **(c)** output conductance $g_{ds}$. Maximum percentage errors are 0.6%, 1.5% and 20%, respectively. Output conductance has larger error at high voltage biases ("A" in **Fig. 7**).

Finally, as evidenced in **Fig. 9**, our trained model and its higher order derivatives satisfy Gummel symmetry tests proposed in [13]. This is unsurprising because ε is symmetric by design, as earlier discussed.

## 7. Summary and Future Work

In this work, a general framework for using deep-learning to assist physics-based transistor modeling is proposed. Using a neural network with roughly 30 neurons, the trained model can predict the simulated IV for a 14nm FinFET technology to within ~1% accuracy. Physics-driven requirements for the neural network are discussed and the model satisfies the Gummel symmetry. This modeling framework can be expanded to account for other transistor parameters such as temperature, device dimensions and *etc*. or be used to predict other transistor characteristics such as capacitance and gate leakage. The neural network can easily be implemented in *VerilogA* for circuit simulation.

Finally, the modeling framework discussed in this work is by no means limited to transistors. It can also be used to model emerging solid-state devices such as light-emitting-diode (LED) or tunneling based transistors (TFET), *etc*, in which only simple, less accuracy equations are available and detailed physics understanding is still in early development. Nevertheless, deep neural network coupled with physics-based model is a very intriguing prospect for modeling next generation semiconductor devices. Work in this vein is dawning and the possibilities are endless.

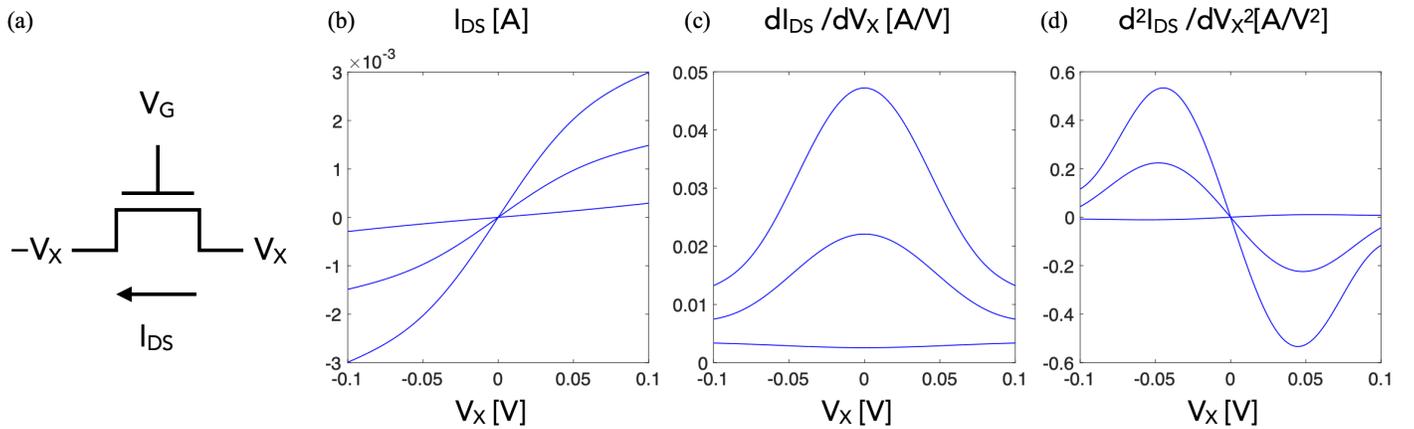

**Fig. 9. (a)** Gummel symmetry test setup. Equal voltages of opposite sign are applied to the source and drain. **(b-d)** $I_{DS}$, first and second derivative of $I_{DS}$ w.r.t. to $V_X$. They are all continuous at $V_X=0$ and pass the symmetry test.

## References
[1] Enz, Christian C., Eric A. Vittoz, and M. O. S. Charge-Based. "transistor modeling: the EKV model for low-power and RF IC design." *John Willey, New York* (2006).
[2] Khandelwal, Sourabh, et al. "BSIM-IMG: A compact model for ultrathin-body SOI MOSFETs with back-gate control." *IEEE Transactions on Electron Devices* 59.8 (2012): 2019-2026.
[3] Gildenblat, G., et al. "PSP Model." *Department of Electrical Engineering, The Pennsylvania State University and Philips Research,(Aug. 2005)* (2005).
[4] Jespers, Paul GA, and Boris Murmann. *Systematic Design of Analog CMOS Circuits*. Cambridge University Press, 2017.
[5] Li, Mingda, et al. "Physics-inspired neural networks for efficient device compact modeling." *IEEE Journal on Exploratory Solid-State Computational Devices and Circuits* 2 (2016): 44-49.
[6] Zhang, Lining, and Mansun Chan. "Artificial neural network design for compact modeling of generic transistors." *Journal of Computational Electronics* 16.3 (2017): 825-832.
[7] Mehta, Kashyap, and Hiu Yung Wong. "Prediction of FinFET Current-Voltage and Capacitance-Voltage Curves using Machine Learning with Autoencoder." *IEEE Electron Device Letters*(2020).
[8] X.Jianjun,and D.E.Root,''Advances in artificial neural network models of active devices,'' in *Proc. IEEE MTT-S Int. Conf. Numer. Electromagn. Multiphys. Modeling Optim. (NEMO)*, Aug. 2015, pp. 1–3.
[9] H. B. Hammouda, M. Mhiri, Z. Gafsi, and K. Besbes, ''Neural-based models of semiconductor devices for SPICE simulator,'' *Amer. J. Appl. Sci.*, vol. 5, no. 4, pp. 785–791, 2008.
[10] W. Fang, and Q.-J. Zhang, ''Knowledge-based neural models for microwave design,'' *IEEE Trans. Microw. Theory Techn.*, vol. 45, no. 12, pp. 2333–2343, Dec. 1997.
[11] Natarajan, S., et al. "A 14nm logic technology featuring 2 nd-generation finfet, air-gapped interconnects, self-aligned double patterning and a 0.0588 μm 2 sram cell size." *2014 IEEE International Electron Devices Meeting*. IEEE, 2014.
[12] Kingma, Diederik P., and Jimmy Ba. "Adam: A method for stochastic optimization." *arXiv preprint arXiv:1412.6980* (2014).
[13] McAndrew, Colin C. "Validation of MOSFET model source–drain symmetry." *IEEE transactions on electron devices* 53.9 (2006): 2202-2206.

## Appendix. Partial derivative of a Neural Network with respect to input

As previously alluded to, accurate prediction for the transconductance and output resistance requires good approximation for the 1st derivatives. To achieve this goal, we first note that the output layer

$$\varepsilon = \sum_{i=1}^{n_L} w_i^L a_i^L + b_i^L \tag{I.1}$$

Differentiate both side with respect to $V_{DS}$ and using chain rule, we get

$$\frac{\partial \varepsilon}{\partial V_{DS}} = \sum_{i=1}^{n_L} w_i^L \frac{\partial a_i^L}{\partial V_{DS}} \tag{I.2}$$

following the standard notation introduced in CS230, the output for the ith neuron in the l-th layer is

$$a_i^l = g(z_i^l) \tag{I.3}$$

Differentiate both side with respect to $V_{DS}$ and using chain rule, we get

$$\frac{\partial a_i^l}{\partial V_{DS}} = g'(z_i^l) \times \frac{\partial z_i^l}{\partial V_{DS}} \tag{I.4}$$

Finally

$$z_i^l = \sum_{i=1}^{n_l} (w_i^l a_i^{l-1}) + b^l \tag{I.5}$$

$$\frac{\partial z_i^l}{\partial V_{DS}} = \sum_{i=1}^{n_l} \left( w_i^l \frac{\partial a_i^{l-1}}{\partial V_{DS}} \right) \tag{I.6}$$

And finally, $T(V_{GS}, V_{GD}) = (V_{GS} + V_{GD}, V_{DS}^2)$

$$\frac{\partial T(V_{GS}, V_{GD})}{\partial V_D} = (-1, 2 \cdot V_{DS}) \tag{I.7}$$

$\partial \varepsilon / \partial V_G$ can be derived in a similar manner. Comparing (I.1), (I.3), (I.5) with (I.2), (I.4), (I.6) together with (I.7). we note that the derivative can be computed by the neural network as shown in **Fig. A1**. Note the similarity between such network and the original network shown in **Fig. 5**. In the derivative network, all the biases are set to zero and the same weights from the original network are copied over. The activation function g(x) is replaced by an element-wise multiplication of g'(x) in which x is fed from the original network. To compute $\partial \varepsilon / \partial V_D$ or $\partial \varepsilon / \partial V_G$, input (0,-1) or (1,1) is feed into the network respectively.

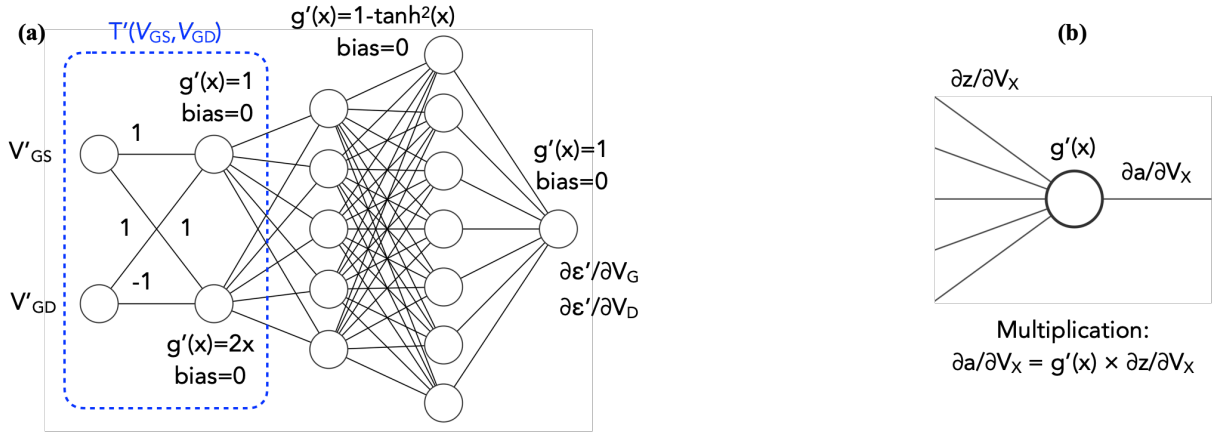

Fig. A1. (a) Neural network to compute the partial derivatives with respect to the input. To compute $\partial \varepsilon' / \partial V_G$, and $\partial \varepsilon' / \partial V_D$, ($V'_{GD}$, $V'_{GS}$) = (1, 1) and (0,-1) are used as the input, respectively. (b) the activation function at each neutron becomes multiplication to the derivative, g'(x) where x and all the weights at each node is fed from the neural network in Fig. 5. All biases are set to zero.